\documentclass[a4paper,10pt,pra,twocolumn]{revtex4}
\usepackage[dvips]{graphics,color}
\usepackage{amsmath}
\usepackage{amsfonts}
\newcommand{\one}{\mathrm{I} \! \! 1}
\usepackage{amssymb}
\usepackage{amstext}
\begin{document}

\title{Entanglement generation from thermal spin states via unitary beam splitters}
\author{D. Markham $^1$}\email{d.markham@imperial.ac.uk}
\author{M. Murao $^2$}
\author{V. Vedral $^1$}
\affiliation{$^1$Optics Section, Blackett Laboratory, Imperial
College, London SW7 2BW, United Kingdom.}
\affiliation{$^2$Department of Physics, Graduate School of
Science,
  University of Tokyo, 7-3-1 Hongo, Bunkyo-ku, Tokyo 113-0033,Japan.}

\begin{abstract}
We suggest a method of generating distillable entanglement form
mixed states unitarily, by utilizing the flexibility of dimension
od occupied Hilbert space. We present a model of a thermal spin
state entering a beam splitter generating entanglement. It is the
truncation of the state that allows for entanglement generation.
The output entanglement is investigated for different temperatures
and it is found that more randomness - in the form of higher
temperature - is better for this set up.
\end{abstract}

\date{\today}
\maketitle

\section{INTRODUCTION}

In quantum information we use quantum mechanics to perform
information tasks in ways exceeding the capabilities of classical
systems \cite{NielsenChuang}. A natural and important question is
then - what is it that gives us the power in quantum information
and where does it come from? It has been shown by Knill and
Laflamme \cite{Knill98} that it is possible to make computations
with exponential speed up over classical algorithms for certain
tasks, using only one pure qubit and a resource of entirely mixed
states. Bose {\it et al.} \cite{Bose01} have proposed a scheme
where only one pure state is needed to generate entanglement
between an atom in a pure state and a thermal field - no matter
what the temperature. Concerning mixed states and nonlocality, in
\cite{Filip02} Filip {\it et al.} used mixed states to violate
Bell's inequalities.

In optics, for example, we know that in the infinite dimensional
case of the harmonic oscillator, if a state can be described as a
statistical mixture of Glauber states, this state cannot be used
to generate entanglement using a beam splitter
\cite{Kim02,Xiang-bin02}. A true maximally mixed state cannot be
used to generate entanglement via any unitary transformation
whatsoever. This is obvious since if the input to an unitary is
proportional to identity, so must its output be, in which case the
state remains unchanged.

Here we wish to address in part the question of how much mixedness
we can deal with and still get entanglement, and how we can do
this. We use the simple idea of change in the dimension of
occupied space to give an example of how we can get around even
maximally mixed inputs to generate entanglement. By expanding the
occupied Hilbert space, the overall entropy remains the same, and
the state is no longer a maximally mixed state for that space, and
thus may have some entanglement. Now, such an operation is not
unitary, in the sense of the input Hilbert space. However, if the
input state is viewed as a truncated state of a higher-dimensional
space on which the unitary acts, and the unitary in question can
expand the occupied Hilbert space, we can get unitary generation
of entanglement.

To illustrate this we use the model of finite truncations of a
thermal state of optical modes entering a beam splitter. Such a
state, truncated to dimension $d$, is exactly analagous to a
thermal spin state, with spin $S=(d-1)/2$. We refer to these as
thermal spin states. Indeed, in principle, we are not restricted
to optics, and can imagine using real spin systems (please see
Sec. \ref{sec: discussion}). The beam splitter is the example we
use of a unitary transformation which increases occupied Hilbert
space. It is exactly this, along with truncation, that allows the
generation of entanglement. In the case of infinite temperatures
this gives unitary entanglement generation from a truncated
maximally mixed state.

We note that one must be careful with considering physical
applications of this model. It is difficult to imagine a beam
splitter type transformation acting on a finite-spin system, and
even then, on the restricted space of the input, this is not a
unitary transformation. In some cases infinite-dimensional systems
can be modelled by considering finite spaces, but this is not what
we are interested in here, since it is the truncation of the space
that leads to the stark difference between the finite and infinite
cases. We present this model simply as in illustration of the
ideas using a real unitary that expands the occupied Hilbert
space. The problems and possibilities of physical implementations
of this model will be looked at in Sec. \ref{sec: discussion}.

We begin in the first section by describing the beam splitter
transformation and defining the measure for entanglement we use.
Next we look at the entanglement generation for a truncated
maximally mixed state. We then look at thermal states, again in
the truncated sense. Finally we show a simple, if inefficient,
protocol proving that the entanglement is distillable for all
states considered, then end with discussion and conclusions.

\section{background}

In optics, the beam splitter is defined by its unitary action on a
two mode Fock state, $|m,n\rangle$, returning the output state
\cite{Compos89},
\begin{eqnarray} \label{eq: Ubs}
U_{bs}|m,n\rangle \equiv &&
\sum_{M=0}^{(m+n)}f(m,n,M)|M,m+n-M\rangle,
\end{eqnarray}
where {\small
\begin{eqnarray} \label{eq: Ubs_2}
f(m,n,M) &=& \sum_{p=max(0,M-n)}^{min(m,M)}\frac{n!m!\sqrt{M!(m+n-M)!}}{p!(p-M+n)!(m-p)!(M-p)!}\nonumber\\
&&\times
T^{p}\bar{T}^{(p-M+n)}R^{(M-p)}\bar{R}^{(m-p)}(-1)^{(M-p)},
\end{eqnarray}}

\noindent and where $T$ and $R$ are the complex transition and
reflection coefficients respectively, with normalisation $|T|^2 +
|R|^2 = 1$. We use this as our definition of the beam splitter on
the general number state (i.e. not restricted to optics).

We notice that for any input state whose density matrix is
diagonal in the two mode Fock basis, the output entanglement is
not effected by the phases of $R$ and $T$, since it can be
absorbed into a phase on the basis states. In all our states this
is the case, so we do not need to consider the phase, and assume,
without loss of generality that $R$ and $T$ are real.

As a measure of entanglement we take the logarithmic negativity
$E_\mathcal{N}$, defined for a given state $\rho$ as,
\begin{eqnarray} \label{eq: logneg}
    E_\mathcal{N}(\rho) &\equiv& \log_2||\rho^{T_A}||_1 \nonumber \\
            &=& \log_2\sum_{i}\left|\mu_i\right|,
\end{eqnarray}
where $\mu_{i}$ are the eigenvalues of $\rho^{T_A}$, the partial
transpose of $\rho$ in subspace $A$. For a state on a Hilbert
space $\mathcal{H}=\mathcal{H}_A\otimes\mathcal{H}_B$, the partial
transpose in subspace A is defined by its matrix elements
\begin{eqnarray}\label{eq: PartialTran}
\langle i_A,j_B|\rho^{T_A}|k_A,l_B\rangle\equiv\langle
k_A,j_B|\rho|i_A,l_B\rangle,
\end{eqnarray}
where
$|i_A,j_B\rangle\equiv|i\rangle_A\otimes|j\rangle_B\in\mathcal{H}_A\otimes\mathcal{H}_B$
form an orthonormal product basis. The logarithmic negativity is a
widely used measure of entanglement \cite{Lee00,Vidal02} and is an
entanglement monotone \cite{Eisert}. We note that a positive value
of this measure does not necessarily imply the existence of
useful, i.e. distillable entanglement (however, for our cases we
prove distillability by independent means).

\section{Maximally Mixed (Thermal) States}

We are now ready to consider unentangled, maximally mixed states,
input to our beam splitter,
\begin{eqnarray} \label{eq: MixIn}
\rho_{in} = \frac{\one}{d}\otimes\frac{\one}{d} =
\frac{1}{d^2}\sum_{m=0}^{2S}\sum_{n=0}^{2S}|m,n\rangle\langle
m,n|,
\end{eqnarray}
with $d=2S+1$. This corresponds to a thermal state at infinite
temperature. The output is then,
\begin{eqnarray} \label{eq: MixOut}
\rho_{out} &=& U_{bs}\rho_{in}U_{bs}^\dag \nonumber \\
&=& \frac{1}{d^2}\sum_{m=0}^{2S}\sum_{n=0}^{2S}\sum_{M=0}^{(m+n)}\sum_{M'=0}^{(m+n)}f(m,n,M)\overline{f(m,n,M')}\nonumber \\
&& \times |M,m+n-M\rangle\langle M',m+n-M'|.
\end{eqnarray}
\\
For any such input state of finite dimension we get an entangled
output. Thus from ``truncated'' maximally mixed input states we
have unitarily derived entanglement. The reason this is allowed is
that the unitary operation does not act only on the space of the
input; in this sense the beam splitter is not unitary on the
truncated Hilbert space of the input.

\begin{figure}[h]
\rotatebox{0}{\resizebox{!}{6cm}{\includegraphics{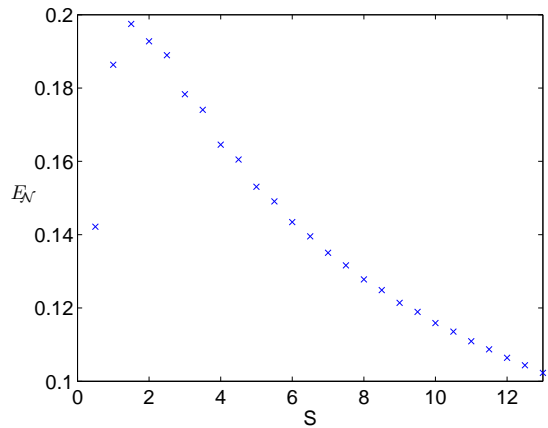}}}
\caption{\label{fg: II}Entanglement of two non entangled,
maximally mixed states passing through beam splitter against S for
$R=T=1/\sqrt{2}$. Here we see an initial peak followed by drop. In
the large $S$ limit this goes to zero - since this is nothing but
a ``classical" state of the harmonic oscillator.}
\end{figure}

Figure \ref{fg: II} is a plot of entanglement against $S$ for
$R=T=1/\sqrt(2)$ for truncated maximally mixed input states. We
see an initial rise followed by a slow tail. We know that in the
infinite $S$ limit the entanglement must fall to zero, since this
is a statistical mixture of Glauber states, which does not
entangle through a beam splitter (such states are also called
``classical" states) \cite{Kim02,Xiang-bin02,Eisert02}.

We may consider the initial rise in entanglement surprising since
in some sense the higher the $S$, the closer we are to a
``classical" state, and so we might think we should get lower
entanglement. The rise can be explained however; as we increase
the size of the spin of the input, the available space for
entanglement increases - the maximal entanglement follows $ln(d)$,
where $d=$ dimension. Then, as more higher levels get occupied,
they begin to destructively interfere and the entanglement begins
to fall with $S$. Similar trends were observed in
\cite{Markham03}, \cite{Arnesen01}, where the entanglement was
studied for spin coherent states sent through a beam splitter and
thermal states in the one-dimensional Heisenberg Model
respectively.

\begin{figure}[h]
\rotatebox{0}{\resizebox{!}{6cm}{\includegraphics{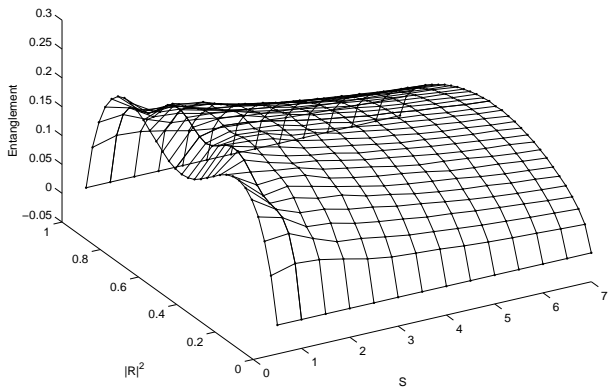}}}
\caption{\label{fg: IIEPR}Entanglement of two maximally mixed
states passing through two arms of a beam splitter against $S$ and
$|R|^2$. We see that the maximal depends on $S$ and is at
$|R|^2=1/2$ for large $S$. The same trend is seen for all
different temperatures of inputs - as long as their are two input
beams, i.e. neither are vacuum (if this is the case, there is only
one maxima at $|R|^2=1/2$ for all $S$). The lines are present
simply to help show the texture of the plot.}
\end{figure}

We can also consider optimality of entanglement generation in
terms of the beam splitter reflectivity $R$ (since we are deling
only with real values, $T$ is set by $R$ through $|T|^2 = 1 -
|R|^2$). Figure \ref{fg: IIEPR} shows the entanglement generated
by maximally mixed input states against $S$ and $|R|^2$. For low
$S$ there are two maxima, at $S=1/2$ they are around $0.2$. As $S$
increases the peaks get closer and for $S$ larger than around $3$
they merge into one at $R=1/\sqrt{2}$. Note, however, that even if
we chose $R$ for each $S$ to give the maximum entanglement, we
will still see an initial peak in entanglement as we go from
$S=1/2$ up.

\section{Thermal States of Different Temperatures}

We now consider a general thermal state of our truncated system.
We assume the Hamiltonian
\begin{eqnarray} \label{eq: hamiltonian}
  H= \sum_{n=0}^{2S} n \hbar \omega|n \rangle \langle n|,
\end{eqnarray}
giving the equalibrium thermal density matrix
\begin{eqnarray}
  \sigma_T = \frac{\sum_{n=0}^{2S}e^{-n \hbar \omega /KT}|n\rangle \langle n|}{Z},
\end{eqnarray}
\\
where $Z=\sum_{n=0}^{2S} e^{-n\hbar\omega/KT}$ is the partition
function.

For two such thermal states incident on a beam splitter at
different temperatures,

\begin{eqnarray} \label{eq: ThermIn}
\rho_{in}&=&\sigma_{T_1}\otimes\sigma_{T_2}\nonumber\\
&=&\frac{1}{Z_1Z_2}\sum_{m=0}^{2S}\sum_{n=0}^{2S}
e^{(-\frac{m}{KT_1}-\frac{n}{KT_2})\hbar \omega}|m,n\rangle\langle
m,n|,
\end{eqnarray}
where $Z_i = \sum_{n=0}^{2S}e^{-n\hbar\omega/KT_i}$, the output
state after the beam splitter transformation is then,
\begin{eqnarray} \label{eq: ThermOut}
\rho_{out} &=& U_{bs}\rho_{in}U_{bs}^\dag \nonumber\\
&=&\frac{1}{Z_1Z_2}\sum_{m=0}^{2S}\sum_{n=0}^{2S}e^{(-\frac{m}{KT_1}-\frac{n}{KT_2})\hbar \omega}\nonumber\\
&&\sum_{M=0}^{(m+n)}\sum_{M'=0}^{(m+n)}f(m,n,M)\overline{f(m,n,M')}\nonumber\\
&&\times |M,m+n-M\rangle\langle M',m+n-M'|.
\end{eqnarray}

\begin{figure}
\rotatebox{0}{\resizebox{!}{6cm}{\includegraphics{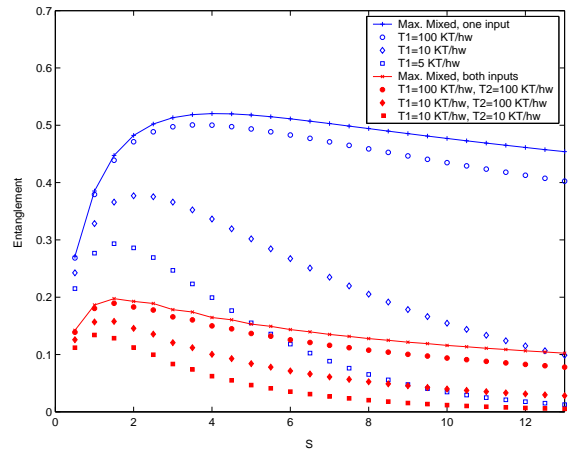}}}
\caption{\label{fg: ALL}Entanglement of various thermal states
against $S$ of the input states for $R=T=1/\sqrt{2}$. The top
four, unfilled, data sets correspond to when one port has a
thermal state entering it and the other only the vacuum. The
bottom four, filled, data sets are when both ports have thermal
states entering them.}
\end{figure}

Figure \ref{fg: ALL} shows some plots for various different input
temperatures, both for the case of a thermal state entering one
port and the other remaining empty (i.e. vacuum state, we can also
think of this as zero temperature), and for thermal states
entering both ports. We see that the general trends are the same
for all temperatures, that of an initial peak followed by a slow
decline. All plots are for reflectivity $R=T=1/\sqrt{2}$. This is
chosen since, when plots similar to figure \ref{fg: IIEPR} were
done for the different temperature inputs, the same trends were
found, thus the overall trends are not affected.

In general the case where the vacuum enters one port gives more
entanglement - this can be explained as resulting from the fact
the two beams do not interfere with one another, which destroys
entanglement.

Starting from the maximally mixed state, which represents an
infinite temperature thermal state, the height of the peak is less
for lower temperatures and the peak occurs earlier. We might
expect the entanglement to be greater for lower temperatures, and
that the maximally mixed state gives the lowest entanglement,
since these are the most disordered and classical like states. The
observed trend can be explained by noticing that for lower
temperatures the higher dimensional states are not as populated,
restricting the possible entanglement. For the zero temperature we
have the ground state which offers no entanglement. We also see a
sharper fall after the peak for lower temperatures.

\section{Discussion} \label{sec: discussion}

For all the output states mentioned here we can devise a protocol
to distill a minimal amount of entanglement. From equations
(\ref{eq: ThermOut}) and (\ref{eq: MixOut}) we can see that if
local projections are made on both arms, onto the subspace spanned
by the states $|0\rangle \langle 0|$ and $|4S\rangle \langle 4S|$,
the remaining state is an entangled state of the form $\rho =
F|0,0\rangle \langle 0,0| + (1-F)(|0,4S\rangle + |4S,0\rangle)
(\langle 0,4S| + \langle 4S,0|)$, which has entanglement for any
finite F.  For finite $S$, $F$ is also finite (hence we have
entanglement). This can be considered as a two dimensional state
and it can be easily shown that it has negative partial transpose,
which, for a two dimensional, bipartite state implies distillable
entanglement \cite{Horodecki97}. As $S$ goes to infinity, F goes
to one and no entanglement can be found in this way as we expect.
All measurements where the projection falls onto the remaining
space are discarded. This scheme shows that for any finite $S$ we
indeed have distillable entanglement for all the thermal states,
though it is not necessarily inefficient and it destroys at least
some, of not most entanglement.

Normally one might assume that from a maximally mixed state,
$\one/d \otimes \one/d$, no entanglement can be extracted via
unitary operations. We see here that this is not the case and that
when considering finite dimensions, there is an extra subtlety
allowed by expanding of contracting the occupied Hilbert space. In
some sense we can consider the operation as expanding the
``mixedness" of the inputs into two, individually less mixed
outputs with some entanglement. More precisely we have the input
output entropy relation $S(\rho_{in1}) + S(\rho_{in2}) \leq
S(\rho_{out1}) + S(\rho_{out2})$, where the inequality comes from
the entropy that has gone to create entanglement (the overall
entropy is conserved since the operation is unitary). From this
perspective it is not so surprising that more entanglement is
generated for the more mixed (higher temperature) thermal input
states. Conversely we can think of the inverse operation as
contracting all the mixedness into a separable mixed state of
lower dimension. This must be true of many unitaries other than
the beam splitter, indeed we may consider the power of a unitary
in terms of expanding or contracting the occupied Hilbert spaces.
For example we can imagine easily a unitary in a six dimensional
space acting on maximally mixed state of two qubits to create
entanglement. The beam splitter simply represents a simple, well
known and physical example of this type of unitary operation. The
scheme by Browne {\it et al.} \cite{Browne03}, represents an
application of a beam splitter used to extend the accessed Hilbert
space to distill gaussian states.

The morel we use here, as mentioned in the introduction, is
seemingly not readily very implementable. This model is intended
as an illustration of the ideas using a real physical unitary that
expands the occupied Hilbert space, rather than a proposal for an
experiment. However, it is hoped that that it could be useful as a
basis for more sophisticated models.

For this, one can think of the input being either a truncation of
an infinite space, or a real thermal state of a finite system.
With infinite dimension, for these thermal states to be physical,
such a truncation should be physically imposed - this could happen
for example, for a finite number of photons in a mode, or atoms in
a Bose-Einstein condensate (BEC). A BEC gas with attractive
interactions, for example, has a natural limit on the number of
particles, imposed by the interactions \cite{Sackett01}. The beam
splitter interaction also exists for BEC \cite{Castin97}. A finite
maximally mixed state for example, could be envisaged as one of a
pair of finite atom-number BECs in a maximally entangled state. In
principle, then, this could offer an implementation of the scheme.

One naive possibility may be the projection of a real optical
thermal state onto a finite-dimensional subspace and then passing
through a beam splitter, giving entanglement. In such a case, the
problem of entanglement generation becomes one of truncation or
projection. This is related to the idea of measurement-induced
entanglement (see e.g. \cite{Lukin03,Sorensen}). In these schemes
a measurement is made at the end of the protocol which gives
guaranteed entanglement conditioned on certain outcomes - it is
the measurement that generates entanglement (whereas we would
measure first and generate entanglement after). These schemes too
implement projection, which does reduce the size of the occupied
space. The difference is that such schemes are concerned with
guaranteed entanglement generation, rather than exploiting the
dimension of the occupied Hilbert space to deal with mixed state
inputs.

Another possibility might be to map a mixed state of some finite
dimension to a higher-dimensional space, and perform a unitary
analogous to the beam splitter in expanding the occupied Hilbert
space, creating entanglement. This could be, for example, a spin
system in a thermal state mapped to the photon number states in a
cavity, then passed through a beam splitter. Feasible mapping of
spin systems onto light has been studied recently (e.g.
\cite{Lukin03,Hammere04r}). In this sense we could begin with a
genuine thermal spin state (even infinite temperature) and extract
entanglement via unitary transformation.

\section{Conclusion}

In this paper we have given an example of distillable entanglement
generation from a maximally mixed state via unitary operation.
This result is explained by the fact that the state is in fact a
truncated maximally entangled state, or equivalently, that the
operation of the beam splitter is not unitary in the finite
dimensional Hilbert space of the input. We have then looked at the
entanglement generation for different thermal state inputs. Again
against intuition, we found that the higher the input temperature,
the more the resulting entanglement. This was put down to the fact
that for lower temperature the population of the higher
dimensional states is lower and hence the possibility for
entanglement generation reduced.

The source of the entanglement, or ``quantumness" for this
scenario can either be thought of as the truncation of the input
state or the power of the unitary to expand the occupied Hilbert
space. There are many natural questions which arise, such as the
minimal unitary way to extract entanglement from a maximally mixed
state (as mentioned, the beam splitter is probably not the optimal
unitary). More comparisons of these finite systems to optics
should lead to more understanding and intuition. As experiments
get closer to high dimensional, finite systems, (e.g.
\cite{Gisin03}), these questions become more and more valid and
useful.

\acknowledgements
This work was sponsored by Engineering and
Physical Sciences Research Council, the European Community, the
Elsag spa, the Hewlett-Packard company and the Asahi Glass
Foundation (Natural Science Research Assistance).

\bibliographystyle{unsrt}

\bibliography{GenBib}

\end{document}